\documentclass[referee,a4paper,12pt,traditabstract]{swsc}

\usepackage{graphicx}
\usepackage{txfonts}
\usepackage{subfigure}
\usepackage{epstopdf}
\usepackage[displaymath,mathlines]{lineno}
\usepackage[authoryear,round]{natbib}
\usepackage[backref]{hyperref}
\usepackage{url}
\usepackage{booktabs}
\usepackage{todonotes}

\hypersetup{colorlinks=true,citecolor=cyan,urlcolor=cyan,linkcolor=blue}

\begin{document}


   \title{Timing of the solar wind propagation delay between L1 and Earth based on machine learning}

   \titlerunning{Solar wind propagation delay L1 to Earth}

   \authorrunning{Baumann and McCloskey}

   \author{C. Baumann
          \inst{1}
          \and
          A. E. McCloskey\inst{1}
          }
   \institute{Deutsches Zentrum f{\"u}r Luft- und Raumfahrt, Institut f{\"u}r Solar-Terrestrische Physik,
              Kalkhorstweg 53, D-17235 Neustrelitz\\
              \email{\href{mailto:carsten.baumann@dlr.de}{Carsten.Baumann@dlr.de}}
             }

   \date{\today}


\abstract
{
Erroneous GNSS positioning, failures in spacecraft operations and power outages due to geomagnetically induced currents  are severe threats originating from space weather. Having knowledge of potential impacts on modern society in advance is key for many end-user applications. This covers not only the timing of severe geomagnetic storms but also predictions of substorm onsets at polar latitudes. In this study we aim at contributing to the timing problem of space weather impacts and propose a new method to predict the solar wind propagation delay between Lagrangian point L1 and the Earth based on machine learning, specifically decision tree models.

The propagation delay is measured from the identification of interplanetary discontinuities detected by the Advanced Composition Explorer (ACE) and their subsequent sudden commencements in the magnetosphere recorded by ground-based magnetometers. A database of the propagation delay has been constructed on this principle including 380 interplanetary shocks with data ranging from 1998 to 2018. The feature set of the machine learning approach consists of six features, namely the three components of each the solar wind speed and position of ACE around L1. The performance assessment of the machine learning model is examined on the basis of a 10-fold cross-validation.

The machine learning results are compared to physics-based models, i.e., the flat propagation delay and the more sophisticated method based on the normal vector of solar wind discontinuities (vector delay).
After hyperparameter optimization, the trained gradient boosting (GB) model is the best machine learning model among the tested ones. The GB model achieves an RMSE of 4.5 min with respect to the measured solar wind propagation delay and also outperforms the physical flat and vector delay models by 50\,\% and 15\,\% respectively. To increase the confidence in the predictions of the trained GB model, we perform a operational validation, provide drop-column feature importance and analyse the feature impact on the model output with Shapley values.

The major advantage of the machine learning approach is its simplicity when it comes to its application. After training, values for the solar wind speed and spacecraft position from only one datapoint have to be fed into the algorithm for a good prediction.

   }        

   \keywords{solar wind propagation --
                ACE --
                machine learning
               }

\maketitle

\section{Introduction}\label{introduction}

Modern society is becoming increasingly vulnerable to space weather impacts. Orbiting satellites for communication and navigation, the once again emerging human space flight and power grids affected by induced currents require timely information on imminent severe space weather events.
One of the main drivers of space weather at Earth is the continuous flow of solar wind.
However, the nature of the solar wind is variable and ranges from a slight breeze of electrons and protons, to fast storms of energetic particles containing ions as heavy as iron.
For the surveillance of the solar wind, several spacecraft have been installed at the Lagrangian point L1. The ACE spacecraft has been, and still is, a backbone for early warnings of severe solar wind conditions \citep{Stone1998}.

For precise forecasts of the ionospheric and thermospheric state, the expected arrival time of these severe solar wind conditions at Earth's magnetosphere is needed. For that purpose, modelling the propagation delay of the solar wind from spacecraft at L1  to Earth has been a long-standing field of research \citep[e.g.][]{Ridley2000,Wu2005,Mailyan2008,Pulkkinen2009,Haaland2010,Cash2016,Cameron2016}. In particular, communication and navigation service users are interested in timely and reliable information about whether to expect a service malfunction or outage at a specific upcoming moment. On the other hand, research topics such  as the timing of the onset of polar substorms \citep[e.g.][]{Baker2002} also benefit from precise information when a potential triggering solar wind feature reaches the magnetosphere.

The above mentioned techniques for the  prediction of the solar wind propagation delay depend on the presence of a shock in the interplanetary medium and provide a velocity-based time delay. Additional approaches to propagating the solar wind include hydrodynamic modeling \citep[e.g.][]{Komle1986,Haiducek2017,Cameron2019}, which model the physical evolution of the solar wind plasma as it travels to Earth.

The measurement of the solar wind propagation delay is usually done by identifying its  distinct features at spacecraft around L1 and Earth orbiting satellites which temporarily probe the solar wind directly (CLUSTER, MMS, Van Allen Probes). These features can be turnings of the interplanetary magnetic field (IMF) \citep{Ridley2000} and even discontinuities in the solar wind caused by Coronal Mass Ejections (CMEs) or Corotating Interaction Regions (CIR) \citep[e.g][]{Mailyan2008} which are used in this study as well.
The magnetosphere on the other hand is also suited to serve as detector of solar wind features.
Magnetometer stations observe the state of Earth's magnetic field on a continuous basis.
As CMEs and CIRs pass the Earth, they can cause significant disturbance to the magnetosphere, leading to so-called sudden commencements in the magnetic field \citep{Gosling1967,Curto2007}. These sudden commencements are detected by ground-based magnetometers across the globe \citep{Araki1977}, allowing for the timing of the solar wind propagation delay just as well as space-based magnetometers.

This study compiles a database of the solar wind propagation delay based on interplanetary shocks detected at ACE and their sudden commencements (SC) at Earth. The database consists of timestamps of the shock detections at ACE and the following SC detections by groundbased magnetometers. The study by \citet{Cash2016} used the same principle to measure the propagation delay.  The database serves not only as a basis to assess the performance of the physical models of the SW propagation between L1 and Earth, but also as a training set for a novel approach based on machine learning (ML).

NASA's well known OMNI database (\url{https://omniweb.gsfc.nasa.gov/}) applies the method of  \citet{Weimer2008} to provide solar wind propagation delays (i.e. the so called timeshift) for 30 $R_E$ ahead of Earth continuously. \citet{Cash2016} tested the ability of Weimer's method in a realtime application and found that it suffers from caveats introduced by additional assumptions applied to the initially shock based method in order to work with continuous data as well.

This study introduces a machine learning method to predict the solar wind propagation delay. The training dataset is defined in a way that only one datapoint of L1 spacecraft data is needed for input, enhancing its flexibility for the use of continuous data as well and may also enable a potential realtime application in the future.  However, as the database for training is comprised of CME and CIR cases only, the valid generalization to a continuous application of the ML approach remains unresolved. The present work can be seen as a first proof of concept that machine learning is indeed able to predict the solar wind propagation delay.

In recent years there has been an ever-increasing number of studies in the field of space weather that have made use of ML algorithms. More specifically, these ML algorithms have been particularly successful for the purpose of prediction, including the prediction of CME arrival times based on images of the Sun \citep{Liu2018}, solar wind properties \citep{Yang2018}, geomagnetic indices \citep{Zhelavskaya2019} and even predictive classification of (storm) sudden commencements from solar wind data \citep{Smith2020}. For an overview on ML applications for space weather purposes we recommend the review by \citet[][]{Camporeale2019}. The advantage of using an ML-based approach, instead of a solely empirical or physics-based model, is that ML models don't require as many a priori assumptions and are generally less computationally intensive.

The present study investigates the possibility to use ML to predict the solar wind propagation delay and is structured as follows.
Section\,\ref{measbase} describes the measurement technique and database of the solar wind propagation delay used in this study.
Section\,\ref{models} introduces the physical models of SW propagation delay and also the new machine learning approach.
Section\,\ref{results} shows the results of the model comparison and an analysis of the trained ML algorithm.
The discussion of the results is carried out in Sec.\,\ref{discussion}.
In Sec.\,\ref{conclusions} we draw the conclusions from our findings.

\section{Delay measurement and database}\label{measbase}
The following section presents the methods of how the solar wind propagation delay has been measured and gives an overview of the contents of the database.
The database of this study is solely comprised of ACE observations of interplanetary shocks and their subsequent sudden commencements at Earth for the years 1998 until 2018.

The database consists of 380 cases that have been identified by ACE and magnetometers on the Earth's surface. The used ACE level 2 data has been provided by the ACE Science Center at Caltech and consists of a timeseries with 64\,s time resolution.
The times of interplanetary shocks have been identified from shock lists \citep{Jian2006,Oliveira2015} and the website \url{ipshocks.fi} maintained by University of Helsinki. These lists combine ACE, Wind and DSCOVR detections of interplanetary shocks and do not always list data for all three spacecraft. In order to increase the number of shock detections for ACE, we have searched ACE data around times that listed  shocks for Wind or DSCOVR but not for ACE. In case ACE did detect the shocks as well, these ACE detections are then added to the database used in this study.
The most recent interplanetary shocks (post 2016) have been identified by visual inspection of ACE data during high geomagnetic activity.
The authors cannot guarantee the capture of all interplanetary shocks from ACE data within the database presented here.

\begin{figure}[htbp]
  \centering
  \includegraphics[width=0.5\textwidth]{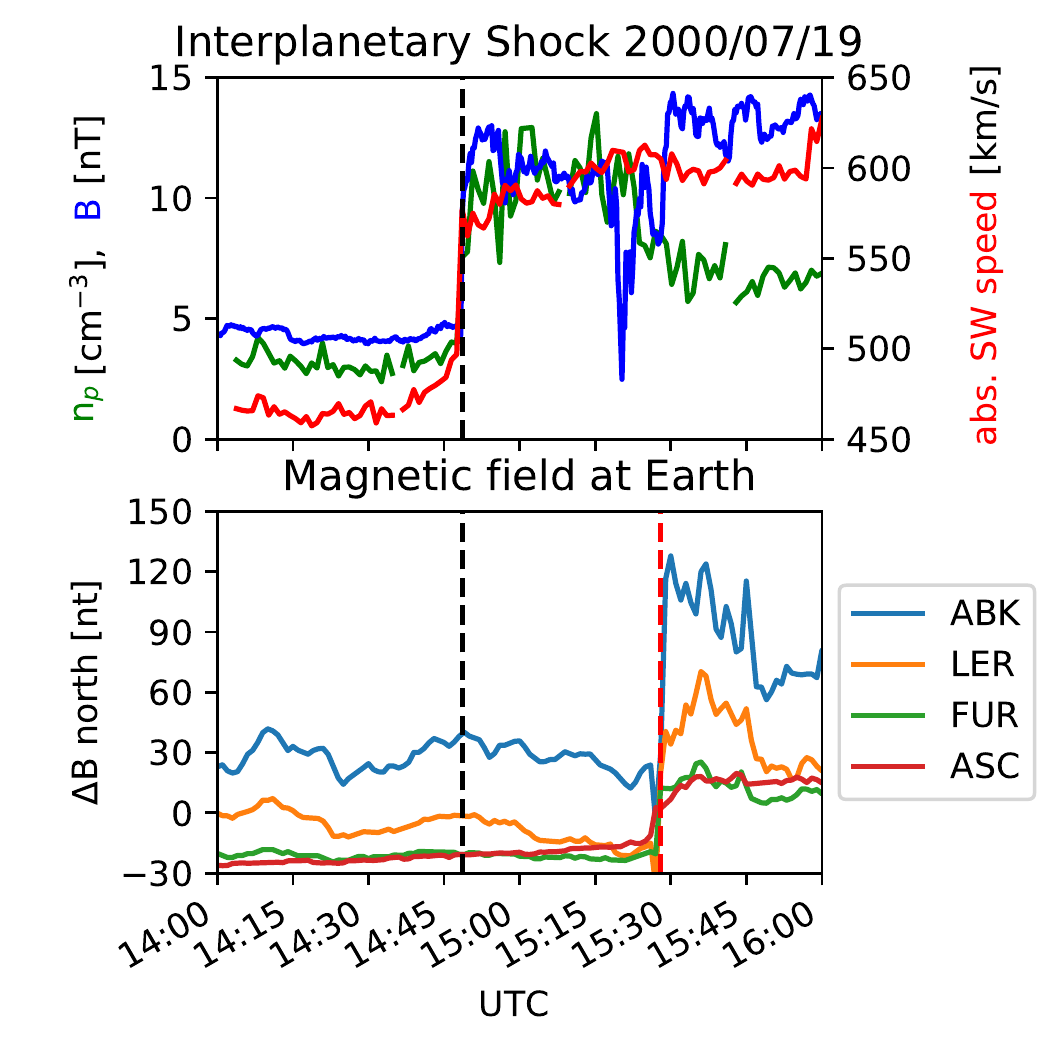}
  \caption{Top panel shows timeseries of proton density $n_p$, magnetic field $B$ and absolute solar wind speed measured by ACE,  bottom panel shows magnetic northward component residuals for magnetometer stations Abisko, Lerwick, Furstenfeldbruck, and Ascension Island, black and red vertical dashed line indicate the interplanetary shock at ACE time 2000/07/19 14:48:35 and a sudden commencement in Earth's magnetosphere at 15:27:00 resulting in a delay of 38 min.}\label{measurement}
\end{figure}

Figure \ref{measurement} (top) shows a typical interplanetary shock as it is measured by the ACE SWEPAM \citep{McComas1998} and MAG \citep{Smith1998} instruments on the 19th July 2000 at 14:48 UTC.
The solar wind propagation delay for this case is identified from the sudden commencement that this interplanetary shock causes in Earth's magnetosphere (Fig. \ref{measurement} bottom panel).
The term sudden commencement describes a magnetospheric phenomena which ground based magnetometers can observe when the magnetosphere is compressed by the impact of an interplanetary shock, i.e. due to the sudden change of the solar wind dynamic pressure \citep[e.g.][]{Curto2007}.
The signature of a sudden commencement is defined as a sudden change of the horizontal component of the magnetic field ($\Delta H$).
$\Delta H$ is the difference between the actual horizontal component H and a baseline ($\Delta H$=$H$-$H_0$).

The sudden commencement is identified from magnetometer data at different locations from high latitudes down to equatorial regions.
The search for the sudden commencements were restricted to times 0.25 to 1.5 h after the detection of an interplanetary shock at ACE.
An additional constraint of the identification of the sudden commencement is that it happens quasi-simultaneously ($<$1 min) at all latitudes \citep[e.g.][]{Engebretson1999,Segarra2015}.
In our study we have used magnetometer stations at Abisko (ABK,  68$^{\circ}$N), Lerwick (LER, 60$^{\circ}$N), F\"urstenfeldbruck (FUR, 48$^{\circ}$N), Bangui (BNG, 4$^{\circ}$N), Ascension Island (ASC, 8$^{\circ}$S), and Mawson (MAW, 67$^{\circ}$S) which are part of the INTERMAGNET consortium \citep{Love2013}.
The identification of sudden commencements is based on 1 min magnetometer data and has been done using the SuperMAG service at JHAPL \citep{Gjerloev2013}. For this analysis, we use the northward component of the magnetic field given by SuperMag to identify the times of sudden commencements.

In ~95\,\% of the cases the identification was unambiguous with the above stations.
However, the identification of sudden commencements during active geomagnetic conditions is not possible at high latitudes.
That is because the already disturbed magnetic field prevents a clear identification of a sudden commencement.
During $\approx$5\,\% of the cases, additional magnetometer stations Tamanrasset (TAM, 22$^{\circ}$N) and Toledo (TOL, 40$^{\circ}$N) have been used to detect the sudden commencement without ambiguities.
An ACE detection/sudden commencement pair was added to the database only in the case of its simultaneous identification at 5 different stations.
Weak interplanetary shocks detected at ACE, that did not cause a geomagnetic sudden commencement, do not allow a measurement of the solar wind propagation delay and are discarded from the database.

The moment of the interplanetary shock at ACE ($T_{ACE}$) has been set to the datapoint when the solar wind speed reaches its downstream (high) value (black dashed vertical line in Fig. \ref{measurement}). So the database consists of just 380 datapoints at the individual time $T_{ACE}$, no timeseries data before the shock is used for the ML propagation delay approach.
The moment of the sudden commencement ($T_{SC}$) at Earth's magnetosphere is set to the sudden increase of $\delta H$.
The propagation delay ($T_D$) is defined as the difference between both times.
\begin{equation}\label{prop_delay}
  T_D=T_{SC}-T_{ACE}
\end{equation}
The systematic error of the measured solar wind propagation delay is at least 2\,min,
because the time resolution of ACE SWEPAM data and of the magnetometer data is 1\,minute each.

\begin{figure}[htbp]
  \centering
  \includegraphics[width=0.6\textwidth]{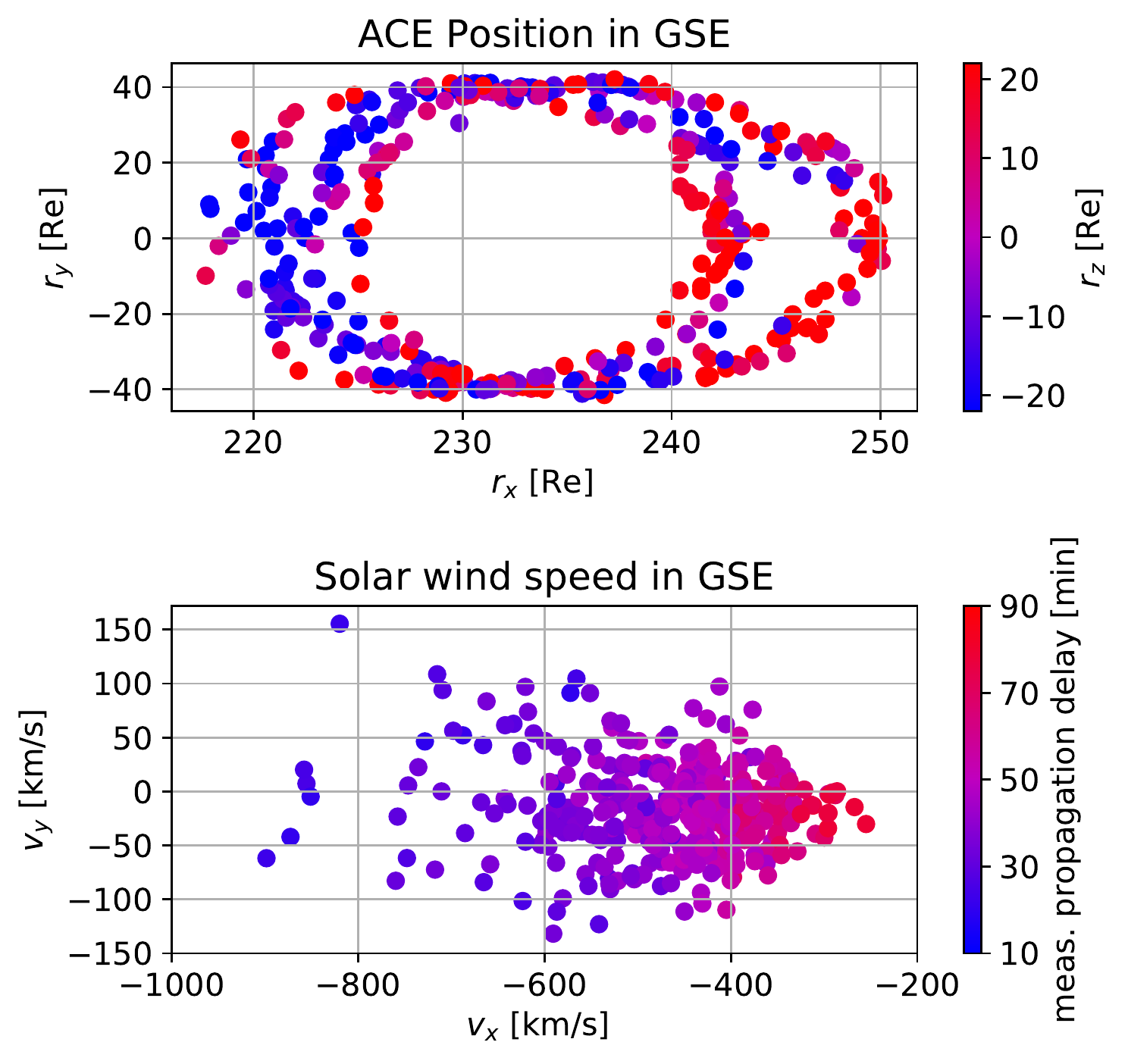}
  \caption{Overview of the database for the calculation of the solar wind propagation delay, top panel shows the ACE position in X, Y, Z, bottom panel shows the measured solar wind propagation delay and the solar wind speed in X, Y, all units are given in GSE coordinates, except the propagation delay [min]. }\label{learnset}
\end{figure}

Figure \ref{learnset} shows the database for the measured propagation delays and additional parameters that have been extracted from the ACE data at the 380 individual times, $T_{ACE}$.
The top panel shows the position of the ACE satellite at the time of the IP shock detection.
It is evident, that the shown positions of ACE represents its Lissajous orbit around Lagrange point L1.
The position of ACE is important for the calculation of the propagation delay and is used for the physical models and the statistical ML model as well.
The bottom panel shows solar wind speed in X and Y-direction measured at $T_{ACE}$ color coded with the measured propagation delay $T_D$.
The propagation delay varies between 20\,min for extremely fast ICMEs around 1000\,km/s and nearly 90\,min for slow shocks around 300\,km/s.
The solar wind speeds and the position of ACE are given in GSE coordinates, therefore higher solar wind speeds show larger negative values in X-direction.
Not shown in Fig. \ref{learnset} is the solar wind speed in Z-direction, which ranges between -150 and 150\,km/s. So the database consists of just 380 datapoints at the individual time $T_{ACE}$, no timeseries data before the shock is used for the ML propagation delay approach.

The database described here has been made available on Zenodo \citep{Baumann2020}. It contains the times $T_{ACE}$ and $T_{SI}$ for all 380 interplanetary shocks and sudden commencements respectively.
It also contains the ACE measurement of all three components of the solar wind speed ($v_x,v_y,v_z$) and the position of ACE ($r_x,r_y,r_z$) at $T_{ACE}$.

\section{Propagation delay models}\label{models}
\begin{figure}[htbp]
  \centering
  \includegraphics[width=0.6\textwidth]{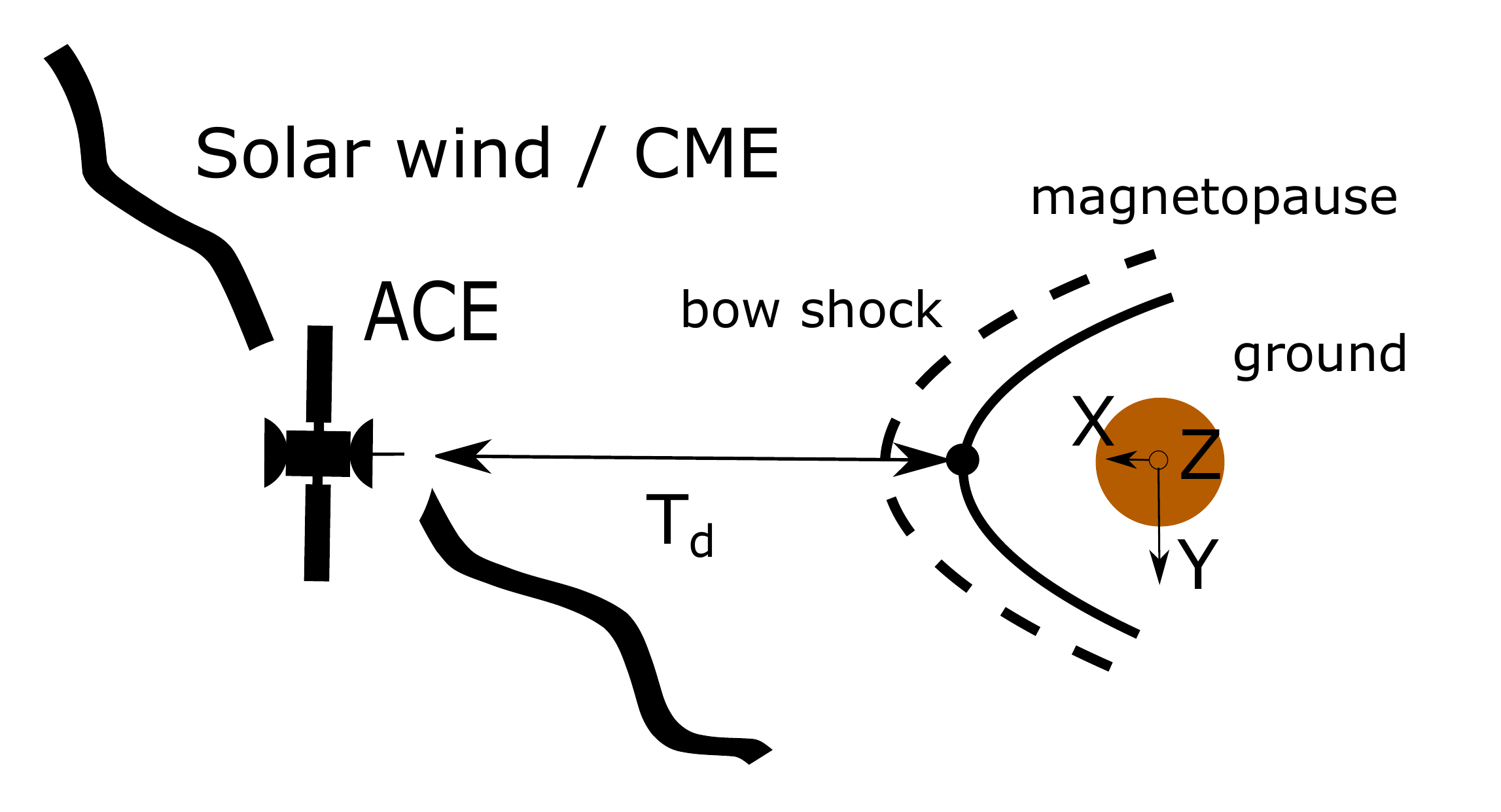}
  \caption{Schematic of the three-part division of the solar wind propagation delay $T_D$ from L1 (ACE) to Earth. Namely, 1. delay between L1 and bow shock, 2. between bow shock and magnetopause, and 3. between magnetopause and ground, after \citep{Mailyan2008}.}\label{schematic}
\end{figure}

The solar wind propagation delay between L1/ACE and Earth can be divided into three parts as indicated in Fig.\,\ref{schematic}.
Firstly, the delay between ACE and Earth's bow shock, i.e. where the solar wind speed drops significantly.
Secondly, the time between the impact at the bow shock and magnetopause.
Thirdly, the delay between the impact at the magnetopause and the start of space weather effects on the ground, e.g. geomagnetically induced currents. While the first part of the propagation delay can be seen as a pure convection of the solar wind \citep[e.g.][]{Mailyan2008}, the other two parts of the delay depend on the geomagnetic conditions and the type of incoming solar wind feature as well as its characteristics.
This study focuses on the delay from the Lagrangian point L1 to the magnetopause (see Sec.\,\ref{measbase}). Timing biases contained within the derived propagation models could account for the differences in timing delay output for the three types of models that will be introduced later in this section.

Otherwise, the solar wind propagation delay is usually addressed by measuring the delay between spacecraft, i.e. a monitoring spacecraft at L1 and an Earth satellite just outside of the terrestrial magnetosphere.
The launch of Wind and ACE for the purpose of solar wind monitoring at the L1 point initiated multiple studies on the physical modelling of the solar wind propagation delay \citep{Ridley2000,Horbury2001,Weimer2003,Mailyan2008}. It has to be noted that the solar wind monitors at L1 are not perfect measures of the solar wind that will eventually interact with Earth's magnetosphere at a later time \citep[e.g.][and references therein]{Borovsky2018}.
Satellites outside of Earth's magnetosphere are able to directly probe the IMF and identify IMF orientation turnings as time stamps for solar wind delay measurements \citep[e.g.][]{Ridley2000}.
However, IMF turnings do not always result in signatures in the magnetosphere which can be used for precise timings of the solar wind propagation delay.
Another method is to use interplanetary shock fronts from CME's and CIRs to identify time stamps at solar wind monitor and detector satellites \citep[e.g.][and references therein]{Mailyan2008}. These shocks are big structures in the solar wind and are unlikely to miss Earth when identified at L1.
This study is based on the propagation of interplanetary shocks from ACE to Earth which are visible as sudden commencements at ground based magnetometers (described in Sec. \ref{measbase}).

There are a number of techniques to model the solar wind propagation delay on a physical basis.  These techniques can be set into two groups.
Firstly, the flat propagation delay which is based on the assumption that the solar wind speed in X-direction is of superior importance for the propagation delay and neglects the other directions.
Secondly, a more sophisticated way to derive the time delay is to use the full three-dimensional space instead.
Here, the vector of solar wind speed, the normal vector of an interplanetary shock front, the position vector of ACE and a target are taken into account.
This method has been termed ``vector delay'' in the following, as it uses the vector representation of the solar wind propagation delay.

This study introduces a new method based on machine learning and compares this method to the above mentioned physical models of solar wind propagation.
In the following, all three methods will be introduced in more detail.

\subsection{Flat delay}

The simplest way to derive the SW propagation delay, from an L1 spacecraft to the Earth's magnetosphere, is to consider the X-direction only. This approach is called 'flat delay' \citep[e.g.][]{Mailyan2008} or was termed 'ballistic propagation' in earlier studies. We have adopted the term flat delay in this study.
The assumption that the solar wind speed is dominated by its X-component is the basis of this approach:

\begin{equation}\label{flat}
  \Delta t_{flat}= \frac{X_{ACE}-X_{T}}{v_x}.
\end{equation}

Here, $v_x$ is the solar wind speed in X-direction, $X_{ACE}$ is the position of ACE along the Earth-Sun line and $X_{T}$ is the target location. In this study we have used a fixed value for the target location just upstream of Earth, i.e., set $X_{T}$ to 15 Earth radii (R$_E$).

The flat delay method has the advantage that it is available as long as there is solar wind speed data from ACE. Its disadvantage is the lack of information on any directionality of the solar wind as well as the interplanetary magnetic field. In addition, the location of ACE around L1 is not fully taken into account.

\subsection{Vector delay}
A more sophisticated approach to model the SW propagation delay uses all available information from ACE, i.e.  full solar wind speed and magnetic field vector.
Also, the position of ACE in all three direction is taken into account.
In the following, this method will be shortly named vector delay. Its derivation is carried out on the basis of the presence of shocks in the interplanetary medium:

\begin{equation}\label{vector}
  \Delta t_{vec}= \frac{(\vec{r_{ACE}}-\vec{r_{T}})\cdot \vec{n}}{\vec{v_{SW}}\cdot\vec{n}}.
\end{equation}

Here, $\vec{r_{ACE}}$ and $\vec{r_{T}}$ are the position vector of ACE and the target location.
$\vec{v_{SW}}$ is the three dimensional solar wind vector and $\vec{n}$ is the normal vector of the interplanetary shock wave heading to Earth.
In this study we use a fixed point of the target location and set $\vec{r_{T}}$ to (15,0,0)\,R$_E$.

The key task of this approach is to determine the normal vector $\vec{n}$ from ACE measurements.
There is a number of techniques available to extract $\vec{n}$ from solar wind speed and magnetic field measurements.
One method is based on the coplanarity assumption, which assumes that the interplanetary shock plane is spanned by two vectors that depend on the magnetic field vector upstream and downstream the interplanetary shock front \citep[e.g.][]{Colburn1966}.
A more sophisticated method applies a variance analysis to define $\vec{n}$  from the minimum variance of the magnetic field  \citep{Sonnerup1967}, maximum variance of the electric field or applying a combination of both \citep{Weimer2003,Weimer2008}.
Other methods even solve the full Rankine-Hugoniot problem of the discontinuity to determine the normal vector \citep{vinas1986fast}.
A good collection and more detail on these methods can be found within the book of \citet{Paschmann1998}.

In this study, we apply the cross product method to derive $\vec{n}$ \citep{Schwartz1998}.
In the following we will shortly recapitulate the underlying derivation.
The coplanarity assumptions allows the definition of $\vec{n}$ from the following cross products.
Magnetic coplanarity (subscript M)  yields the following formula for the normal vector:
\begin{equation}\label{magnetic_cop}
\vec{n_M}=\pm\frac{(\vec{B}_{\,d}\times\vec{B}_{\,u})\times\Delta\vec{B} }{\left| (\vec{B}_{\,d}\times\vec{B}_{\,u})\times\Delta\vec{B}  \right|}
\end{equation}
$\vec{B}$ and $\vec{V}$ are the three dimensional vectors of the magnetic field and solar wind speed measured at ACE.
Vectors with subscript $d$ denote downstream conditions and vectors with subscript $u$ denote upstream conditions.
The $\Delta$ sign indicates the difference between downstream and upstream conditions.

The three following equation rely on the coplanarity of $\vec{n}$ with a mix of magnetic and solar wind vectors (subscript MX1-3):
\begin{equation}
\vec{n_{MX1}}=\pm\frac{(\vec{B}_{\,u}\times\Delta\vec{V})\times\Delta\vec{B} }{\left| (\vec{B}_{\,u}\times\Delta\vec{V})\times\Delta\vec{B}  \right|}
\end{equation}
\begin{equation}
\vec{n_{MX2}}=\pm\frac{(\vec{B}_{\,d}\times\Delta\vec{V})\times\Delta\vec{B} }{\left| (\vec{B}_{\,d}\times\Delta\vec{V})\times\Delta\vec{B}  \right|}
\end{equation}
\begin{equation}
\vec{n_{MX3}}=\pm\frac{(\Delta\vec{B}\times\Delta\vec{V})\times\Delta\vec{B} }{\left| (\Delta\vec{B}\times\Delta\vec{V})\times\Delta\vec{B}  \right|}
\end{equation}
Also the difference between downstream and upstream solar wind speed can be used to derive the normal vector of the interplanetary shock front.
\begin{equation}\label{speed_cop}
\vec{n_V}=\pm\frac{\vec{V_d}-\vec{V_u} }{\left| \vec{V_d}-\vec{V_u}  \right|}
\end{equation}

Upstream and downstream conditions of $\vec{B}$ and $\vec{V}$ have been deduced from averaging measurements  5\,min before (upstream) and after (downstream) the shock.
For further analysis all five cross product methods (Eq. \ref{magnetic_cop}-\ref{speed_cop}) have been evaluated and the mean $\vec{n}$ has been applied to the derivation of the vector delay (Eq. \ref{vector}).

The advantage of the vector delay in comparison to the flat delay is its higher accuracy \citep[e.g.][]{Mailyan2008}.
The major disadvantage of the vector delay is the requirement of a discontinuity (CME or CIR) within the solar wind to derive $\vec{n}$.
This requirement prevents a timely evaluation of the vector delay and makes its application to a realtime service nearly impossible.

\subsection{Machine learning delay}
\label{ml_delay_sec}
The aim of this new machine learning approach for SW propagation delay modeling is to combine the advantages of the flat and vector delay methods. Specifically, the all-time applicability of the flat delay and the higher accuracy of the vector delay. The all-time applicability is achieved by the nature of the used database. The database only consists of a single ACE datapoint downstream for each interplanetary shock and does not include data from the timeseries several minutes  before or after the shock. As a consequence the trained ML model does not know about the presence of a shock front and can be used with continuous data as well. A higher accuracy is expected for the ML approach because the whole position vector of ACE and the Solar wind vector, as similarly applied to the vector delay method, are used for training of the model.

The choice of a machine learning model is often an arbitrary one, mostly dependent upon the computational cost for the specific problem, and in principle many ML algorithms could be applied to the same problem. In this paper we choose to investigate the application of three  different ML models in predicting the solar wind propagation delay, namely Random Forest Regression (RF) \citep{Breiman2001}, Gradient Boosting (GB) \citep{Friedman2001} and Linear Regression (linReg; represented as ordinary least square regression in \citet{scikit-learn}).

RF and GB algorithms generate ensembles (forests) of decision trees to make predictions. The main difference between RF and GB model is the characteristics and evaluation of the decision trees in order to produce an output. The RF model builds independent decision trees and produces its result on the basis of an equally weighted average over all trees, a method called bootstrap aggregation in statistics. The GB algorithm improves the performance of individual trees based on a recursive learning procedure, known as boosting.  The main reasoning for this choice of models was to firstly enable direct comparison between the RF and GB models, to quantify if the use of an ensemble-based ML model make a significant improvement to the overall performance. Additionally, the linReg model was included as a simple benchmark for comparison with all other models.
Both decision tree algorithms exhibit a high degree of versatility and interpretability with regard to the underlying problem, while also demonstrating good overall performance in general \citep[e.g.][and references therein]{Biau2016,Zhang2015}.  Training and testing of the ML algorithms have been carried out using the Scikit-learn Python package \citep{scikit-learn}.

This study uses these machine learning algorithms in their regression representation.
Therefore, the machine learning SW propagation delay can be described as the output from a function with the feature vector \vec{x}:
\begin{equation}
    \Delta t_{ML}=f_{\mathcal{D}}(\vec{x}).
\end{equation}
The notation $f_{\mathcal{D}}$ describes a machine learning algorithm trained on the data set $\mathcal{D}$.
The feature vector $\vec{x}$ contains six features that includes each component of the position vector of ACE ($r_x,r_y, r_z$) and the measured solar wind speed vector ($v_{x}$, $v_y$, $v_z$).
The data set contains overall 380 samples and is described in Sect. \ref{measbase}. Each sample represents an interplanetary shock measured at ACE and detected as sudden commencement in the magnetosphere.
The samples contain the above mentioned feature vector $\vec{x}$ and the measured solar propagation delay which is the target variable ($Y$). To avoid biases between different ML models, $\vec{x}$ is standardized before training.

\subsubsection*{Hyperparameter optimization}
Most machine learning algorithms contain parameters which control their general behavior, the so called hyperparameters. In case of decision tree algorithms, these hyperparameters define the characteristics of the decision trees generated and the number of trees in the forests.

For that purpose Bayesian optimization based on the Gaussian process is often applied \citep[e.g.][and references therein]{Swersky2013}.
This study  also follows this paradigm by using the scikit-optimize \citep{Head2020} python package.
For a hyperparameter optimization the database is split into training, testing, and validation set.
In this case, the validation set contains 10\,\% of the database. The remaining 90\,\% is used for the Bayesian optimization using an internal 5-fold cross-validation. The Bayesian optimization tries to minimize the models root mean square error (RMSE) with respect to the measured SW propagation delay. This is done by consecutively changing the underlying hyper parameters and finding the best set of hyper parameters in this process.

Table \ref{optimized_parameters} shows the hyperparameters that have been taken into account during the optimization of the decision tree models. This table shows also the default parameters used in SciKit-learn. The random forest heavily relies on the number of trees that are generated, as this algorithm reduces its output variance by averaging over many random trees. The gradient boosting methods does not require so many trees but the learning rate is important here as it governs the recursive improvement of individual trees. Other hyperparameters define how the branches of the decision trees are generated, i.e. min samples split, min samples leave, max features per tree and max depth of the decision trees.
For a closer description of these hyperparameters we refer to the SciKit-learn documentation \citep[][scikit-learn 0.22.1]{scikit-learn}.

\begin{table}
  \centering
    \caption{Default and Bayesian optimized parameters for the random forest and gradient boosting algorithms, learning rate is not a hyperparameter of the random forest algorithm, minimum impurity decrease was also optimized but did not show a change from 0.0, variable max depth of the default random forest is defined as the expansion of tree until endpoints contain less than [min\_samples\_split] samples. }\label{optimized_parameters}
  \begin{tabular}{ l l l l l l l }
    \toprule
    Algorithm & \# Trees  &max features&min samples &min samples  &  max  &learning \\
  default/optimized&  &per tree & split& leaf  & depth &rate\\\midrule
    Random Forest & 100/800 & 6/3 &2/2 &1/1 & var./11&N.A.\\
    Gradient Boost & 100/490 & 6/3 &2/20 &1/8 &3/5 &0.1/0.02 \\
    \bottomrule
  \end{tabular}

\end{table}

\begin{figure}[htbp]
  \centering
  \includegraphics[width=0.6\textwidth]{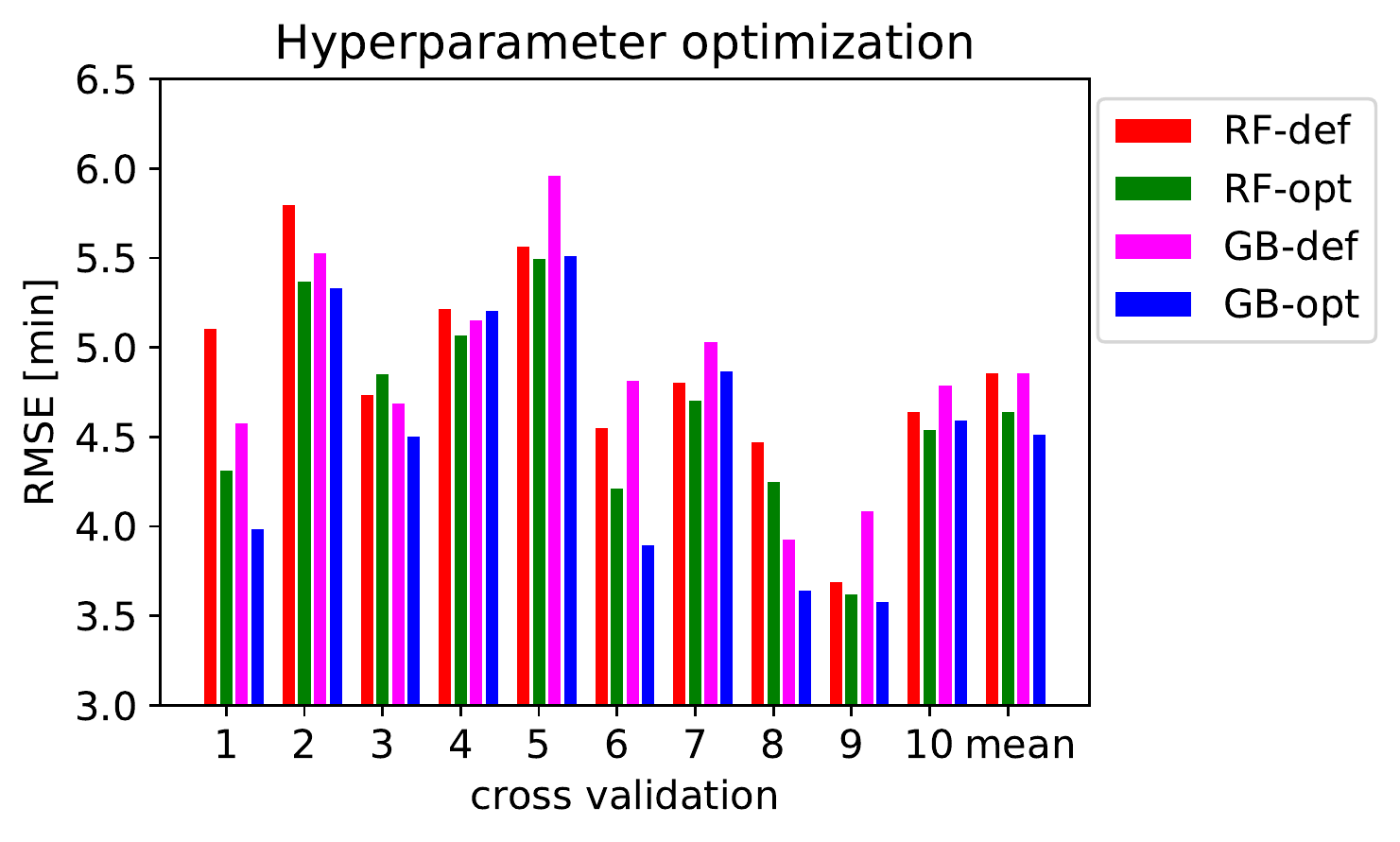}
  \caption{Histogram of RMSE for the 10 fold cross validation of default (red) and optimized (green) random forest as well as the default (magenta) and optimized (blue) gradient boosting algorithm, the fold number is indicated on the x-axis together with the mean of all folds. }\label{hyperparameter}
\end{figure}

The next step is to investigate the performance of the optimized ML algorithm compared to the default algorithm. For that purpose we performed this time a k-fold cross-validation, using 10 folds. Here, the full dataset is split into 10 parts where each segment is iteratively used as the test set with the other remaining segments used to train the model.
To prevent biases due to training data being ordered in time, the database has been randomly shuffled first. The segmentation is then done in a stratified way, so that each fold contains a training data distribution(shock cases) that represents the full range of measured solar wind propagation delays.
The RMSE with respect to the measured SW delay acts as a performance metric. Figure \ref{hyperparameter} shows all ten folds in a set of histograms. The best algorithm to predict the SW delay is the optimized gradient boosting method with a mean RMSE of 4.5\,min, closely followed by the optimized random forest with 4.7\,min. That is an improvement of 8\,\% and 5\,\% for the optimized gradient boosting and random forest algorithm, respectively, compared to their default counterparts. The GB method benefits more from hyperparameter optimization than the random forest, however the RF in its default version performs slightly better than the default GB.  Applying a  cross validation shows that some SW propagation delay cases of the database are more difficult to predict than others. This behavior is also independent of whether the ML algorithms are used with optimized or default hyperparameters.
In the following we will compare the optimized ML results to the flat and vector delay method.

\section{Results}\label{results}
\subsection{Comparison of ML and physical delay models}

At first we investigate the performance of all delay model is investigated for the example case shown in Fig. \ref{measurement}. Table\,\ref{example_case} summarizes the delay model outputs for this case and its actual SW propagation delay measurement of. The ML models were trained with the whole database excluding only the case on 2000/07/19 and used the feature vector (248\,R$_E$, 13\,R$_E$, 18\,R$_E$, -556\,km/s, -76\,km/s, 91\,km/s) for its prediction. The RF  and GB model  as well as the vector method predict the SW propagation delay for this specific CME with less than one minute deviation from the measurement. The linear regression and flat delay method overestimate the delay by 5 respectively 6 min.

\begin{table}
  \centering
    \caption{Solar wind propagation delay predictions using all five methods for the example CME on 2000/07/19  (see Fig.\,\ref{measurement}) with feature vector (248\,R$_E$, 13\,R$_E$, 18\,R$_E$, -556\,km/s, -76\,km/s, 91\,km/s). The two digits after the decimal point are insignificant, as the measurement error is in the order of two minutes.}\label{example_case}
  \begin{tabular}{ l l l l l l l }
    \toprule
 delay model & RFreg  &GBreg& vector & flat &linReg & measured \\\midrule
   SW delay [min]& 37.(85) & 38.(45) &39.(05) &44.(23) &43.(12)&38.(41)\\

    \bottomrule
  \end{tabular}

\end{table}

For an statistical assessment of the ML performance we use the cross-validation approach used in the hyperparameter optimization (see Fig. \ref{hyperparameter}).
Stratified K-fold cross validation is a robust method to investigate the performance of a ML algorithm.
This approach prevents positive bias when interpreting the statistical nature of the ML results compared to other methods.
The comparison contains three ML algorithms, i.e. random forest, gradient boost and linear regression, and the flat and vector method to model the SW propagation delay.
Taking simple linear regression into account allows us to investigate if the more sophisticated ML algorithms achieve greater performance when compared with a very simplistic model.

\begin{figure}[htbp]
  \centering
  \includegraphics[width=0.999\textwidth]{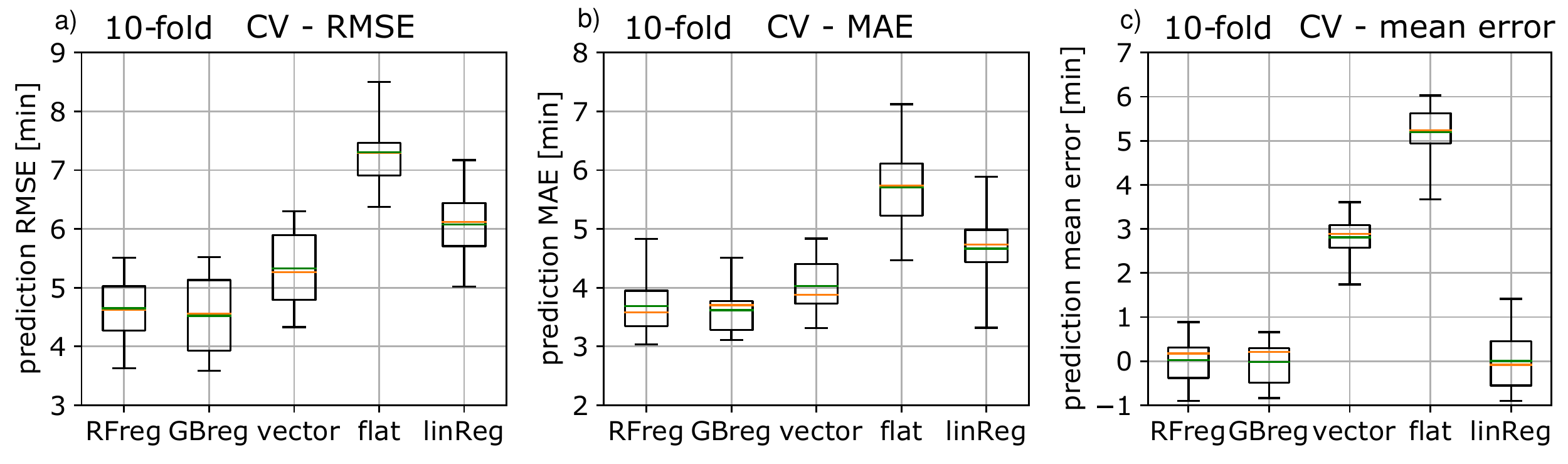}
  \caption{Comparison of model performance  based on a 10-fold cross validation (CV)  between random forest (RFreg), gradient boost (GBreg), vector delay, flat delay, and basic linear regression (linReg) shown as box plot for the performance metrics a) RMSE , b) mean absolute error (MAE) and c) mean error . Each box contains 10 folds from the cross validation, its mean is shown in green, the median in orange, the box edge show the 25/75 percentile and the whiskers indicate the full range. }\label{comparison}
\end{figure}

In order to achieve a reasonable comparison of the trained ML algorithms to the flat and vector methods to model the solar wind propagation delay, we use the same test sets to derive RMSEs for all methods.
Figure \ref{comparison} contains the results of the 10-fold cross-validation for all five methods to model the SW propagation delay.
Here, the metrics (RMSE, MAE and mean error) serve as an accuracy measure for each method's capability to predict the solar wind propagation delay.
The gradient boosting and random forest results are the same as in Fig. \ref{hyperparameter} for the optimized algorithms.
The linear regression model has been trained and tested with the same cross-validation folds.
The performance of the physical models is based on the test sets only.

The comparison using RMSE metric (Fig.\,\ref{comparison} a) reveals that the decision tree models (RF and GB) boosting perform better than  both the LinReg and Fphysical models.
The ten-fold cross-validation shows that RF and GB perform almost the same with a variation between 5.5 and 3.5\,min and a mean of only 4.7 and 4.5 min, respectively.
The best physical method to model the SW delay is the vector method with a mean RMSE of 5.3\,min. However, its range is also higher, ranging from 6.5\,min down to 4.1\,min.
Linear regression performs with a mean RMSE of 6.1\,min  in the fourth place.
The flat method shows the worst result among all studied algorithms and can model the solar wind delay with a mean RMSE of only 7.3\,min.

The same cross validation, but using the mean absolute error as a metric (Fig.\,\ref{comparison} b), show a similar ranking of the different approaches to predict the SW propagation delay. The mean error metric (Fig.\,\ref{comparison} b) shows a different behaviour. All ML models show mean errors around 0 minutes, which is expected from their statistical nature. Only the physical models show a bias when considering the mean error. The vector delay overestimates the SW delay by 2.8 min while the flat delay overestimates by even 5.2 min.

\begin{figure}[htbp]
  \centering
  \includegraphics[width=0.7\textwidth]{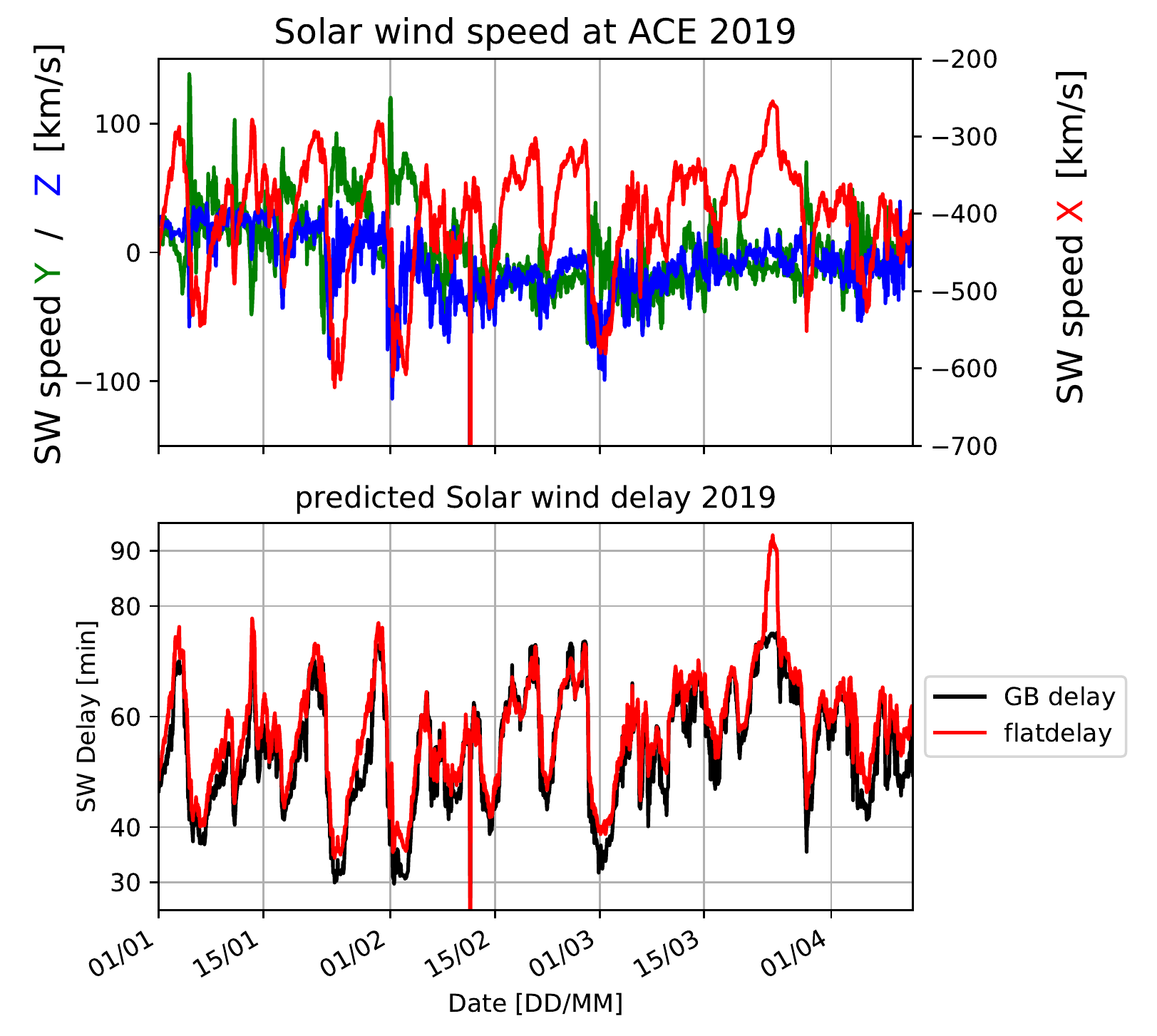}
  \caption{Comparison of SW propagation delay prediction for the beginning of 2019 using the flat delay method and a fully trained GB model.  Continuous ACE level 2 hourly data is used for input. }\label{Flat_ML comparison}
\end{figure}

Figure\,\ref{Flat_ML comparison} shows  a comparison of the flat method and fully trained GB algorithm predictions based on ACE level 2 data from the first 100 days in 2019. Interplanetary shocks from 2019 onward are not part of the training data set, the chosen period is therefore completely unknown to the GB model. This comparison points at the ultimate future application for machine learning based SW propagation delay predictions, i.e. a realtime operational setting. It has to be noted that the analysis in Fig.\ref{Flat_ML comparison} is only qualitative as no ground truth of the solar wind propagation delay is available to the authors. A rigorous validation of this kind of continuous application is subject to a future study.

During the majority of the time period, both predictions are qualitatively similar and vary between 30 and 70\,min.
However, when the solar wind speed in Y and Z-direction is non-zero the trained GB model predicts propagation delays several minutes shorter than the flat method.
Between 15. March and 01. April there is a time period of solar wind speed as low as -250 km/s in the X-direction. While the flat delay predicts propagation delays of up to 90 min, the GB model output stays around 77 minutes. This behavior can be explained by the lack of training data for these very low values of solar wind speed. In the context of this work, there is no evidence that any heliospheric shock with solar wind speed at this low level generates a detectable sudden commencement in the magnetosphere.

The further analysis concentrates on the GB machine learning model only.
The RF as well as the linear regression are discarded from that analysis.

\subsection{Performance Validation}

\begin{figure}[htbp]
  \centering
  \includegraphics[width=.7\textwidth]{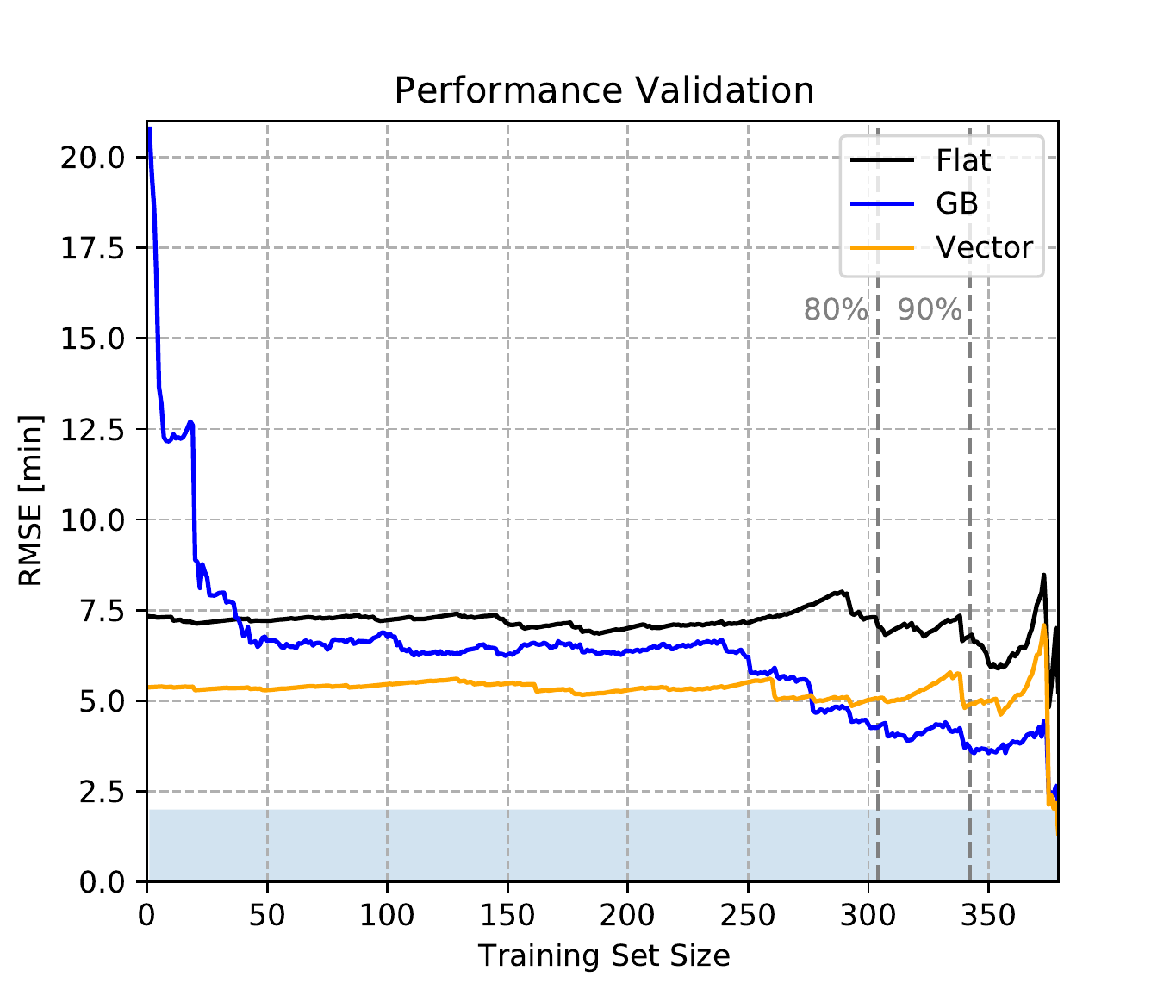}
  \caption{Comparison of model performance (RMSE) with varying train/test set ratios for the gradient boosting (blue) model. Physical models, vector delay (orange) and  flat delay (black), are evaluated on the same test set size which is equal to 380 - training set size.  The shaded region corresponds to the estimated error in the delay measurement.}\label{realtime_rmse}
\end{figure}
For the purpose of implementing these machine-learning models in predicting the solar wind propagation delay, it is important to consider performance validation of the entire dataset. In the field of machine learning it is standard practice to choose a train/test ratio of 80/20 (i.e., 80\,\% of the data is training and 20\,\% is testing) or 90/10 when verifying the performance of models. The choice of these ratios is typically subjective and provides only a single-valued estimate of each model's performance. An analysis of the dependence of the performance metric (RMSE) on the selection of training/testing set ratios is presented in Fig.~\ref{realtime_rmse}. As the flat and vector method are simply analytical models based on physical assumptions, they do not rely on statistical training, evidenced by the near-constant RMSE value for the variable test sets.

As expected, the performance of the GB model depends significantly on training set size and tends to converge toward a stable RMSE when the training set contains at least 40 samples. Already with this small amount of training data, the GB model can predict the SW propagation delay with a lower RMSE than the simple flat method.
Otherwise, relying on these small training set sizes the GB model cannot outperform the vector delay that achieves lower values of RMSE, i.e., better performance.

However, the performance of the GB model begins to increase again when training set size reaches $\approx$\,250 cases of the total data set. Leading to ultimately better performance of the GB model than either the flat or vector methods when using more than 270 cases for training. It is also interesting to note that as the testing set size decreases to $<10\,\%$ of total data set size, all model RMSE values increase, i.e, their performance decreases. This behaviour can be explained by considering the case of low-number statistics, the testing set size is not sufficiently large enough and therefore the RMSE values have increasingly high variance. Hence, performance values in this range are deemed statistically unreliable.

Using the standard 80/20 split case, the flat, vector and GB models achieve RMSE values of 7.1, 5.1 and 4.3\,min respectively. In the case of a 90/10 split, the flat, vector and GB models achieve RMSE values of 6.8, 4.9 and 3.7\,min, respectively. In both cases, the GB model out-performs both the vector and flat methods, a result that was previously reflected in the k-fold cross validation analysis (see Fig.\ref{comparison}).

\subsection{Explaining the gradient boosting results}
\begin{figure}[htbp]
\centering
\includegraphics[width=0.5\textwidth]{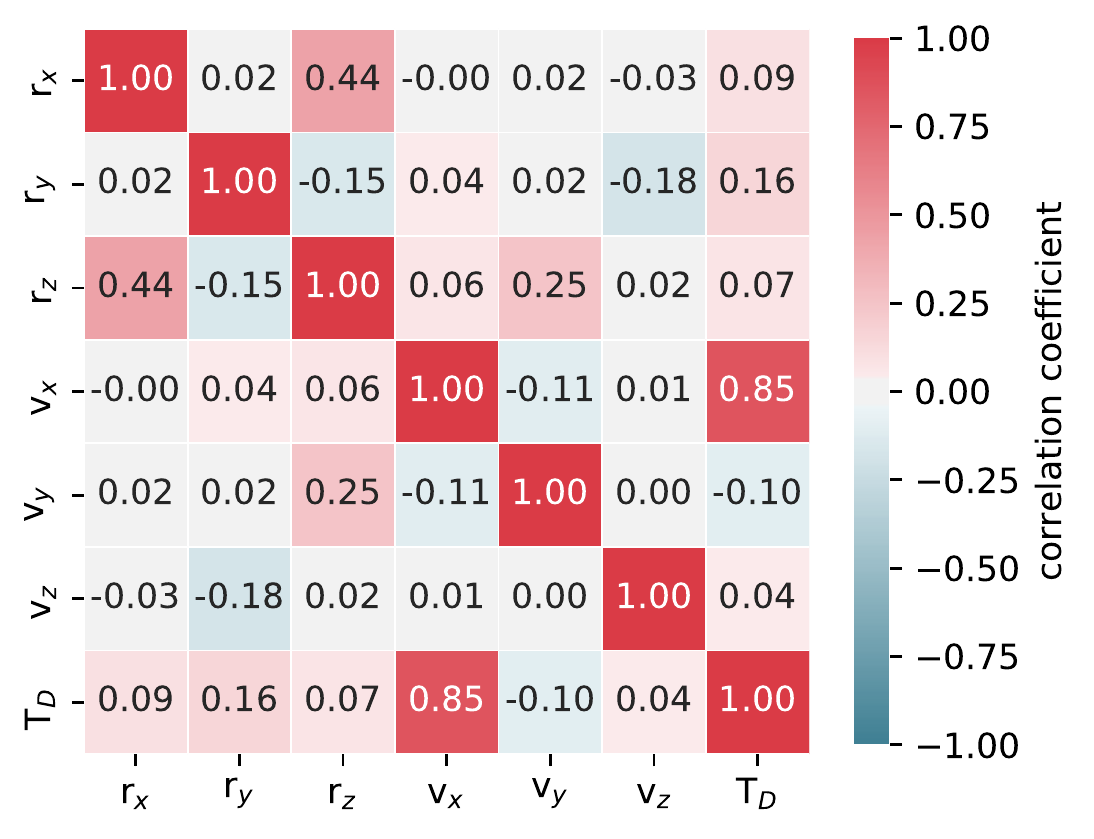}
\caption{Correlation matrix of the machine learning features (position of ACE ($r_x$, $r_y$, $r_z$), solar wind speed ($v_x$, $v_y$, $v_z$)) and the measured solar wind delay $T_D$ of the database.}\label{correlation}
\end{figure}

To improve the understanding of the physical mechanisms, information from the trained gradient boosting algorithm is extracted.
To start with, Fig. \ref{correlation} shows the correlation matrix of the used features based on the underlying database ( cf. \ref{measbase}) and the target value ($T_D$). There are two combinations of enhanced correlation among the features itself.
Firstly, the position of ACE in X and Z-direction have a correlation index of 0.44.
This correlation originates from the nature of the ACE' orbit around L1.
Overall, the dataset is only slightly correlated and all features are expected to contribute to the prediction based on the machine learning model.
In addition, the solar wind speed in X-direction is strongly correlated ($c=0.85$) to the measured solar wind propagation delay, that is expected from the assumption in the flat delay approach. The other features are hardly correlated to the target variable.
The identified correlations has to be taken into account in the further explanation of the trained GB model.

One way of explaining trained machine learning algorithms is to derive the so-called feature importance (FI).
FI is usually derived to identify the subset of features which has the biggest impact on ML model accuracy and robustness.
Selecting only the most valuable and relevant features also decreases the time needed to train ML models.

However, there are many different interpretations of how FI can be retrieved from machine learning algorithms \citep[e.g.][and references therein]{Hapfelmeier2013}.
This study performs drop column FI as this method is able to identify unambiguously the feature importance from random forests \citep{Strobl2007}.
Drop-column FI  determines the change in performance when a feature (column) is left out (dropped) of the feature set to train the GB model when compared to a fully trained model.
As a performance metric the RMSE is used here again.

Drop-column FI values can be positive and also negative. Positive values indicate that leaving out a certain feature increases the RMSE of the ML model.
Features showing a negative FI indicate that leaving out this feature reduces the RMSE of machine learning model, i.e. the performance increases.
The drop column feature importance ($DC_{FI}$) of feature $x$ for a trained ML algorithm can be represented as follows:
\begin{equation}\label{dropcolumn}
DC_{FI}(x)=RMSE(x\notin F)-RMSE(x\in F).
\end{equation}
Here, $RMSE(x\notin F)$ is the RMSE obtained from a trained random forest leaving feature  $x$ out of the used feature set $F$.  $RMSE(x\in F)$ is the RMSE of the fully trained random forest. Both RMSE's are evaluated from the same test dataset.

\begin{figure}[htbp]
  \centering
  \includegraphics[width=0.6\textwidth]{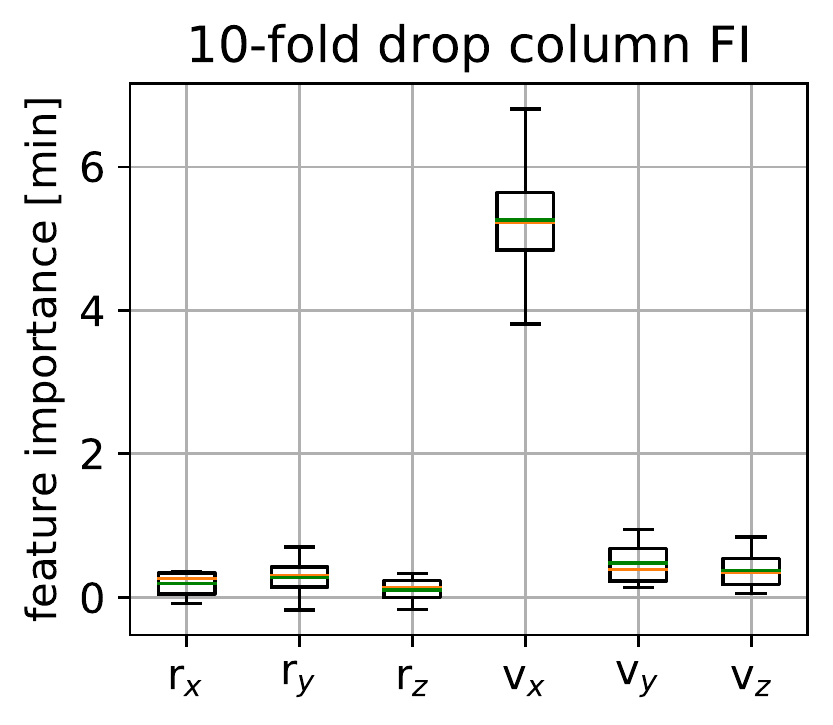}
  \caption{Drop column feature importance of the GB model using 10-fold cross validation, whispers show the full range, the box show the 25\,\%/75\,\% percentiles, the green line indicates the mean, the orange line indicates the median.}\label{importance}
\end{figure}

Figure \ref{importance} shows the results of drop-column FI determination.
To identify the statistical behaviour of the 6 features of the random forest model, a 10-fold cross-validation is performed for the drop column FI.

Each box in  Fig. \ref{importance} contains the  mean importance as well as its variability for each feature.
By far the highest increase in RMSE occurs when the solar wind speed in X-direction $v_x$ is left out of the feature set, FI values range between 3 and 6 min with a mean of 4.6\,min and the median at 4.5 min.
All other features show mean importances below  one minute, some folds within the cross validation show even negative values.
The solar wind speed components in Y and Z-direction($v_y$, $v_z$) show smaller feature importance ($\approx$ 30\,sec), however all train/test folds show positive values.
The FI of the position of ACE is even lower. The FI of the ACE position X and Y-component have positive mean values around 10-20\,s.
For the slightly correlated feature $r_z$ we find an importance close to zero, i.e. $r_z$ does not contribute to the performance of the trained random forest.

Drop-column FI only gives a general view of the trained GB model, but the underlying functioning of the algorithm remain unresolved.
In order to get a glimpse into the GB itself, Shapley values can open a view into its depth.
\citet{Shapley1953value} proposed a measure to identify the bonus due to cooperation within a cooperative game.
The surplus that each player contributes to the outcome of the game is called the Shapley value.
The principle of a cooperative game can also be applied to the GB regression of this study.
Here, the ML features resemble Shapley's cooperative players.
The python package SHAP provides functionalities to derive Shapley values from trained ML algorithms \citep{Lundberg2017} and was also used for the random forest analysis in this study.

A fully trained GB model, i.e, all features and samples have been used for training, is the basis of the Shapley value analysis.
Shapley values can be calculated for each sample of the database (Sect. \ref{measbase})
In this study  the Shapley values represent the changes of the predicted solar wind propagation delay with respect to a mean model output. A negative Shapley values indicate that the model output for this individual sample results in a shorter solar wind delay compared to the mean model output.
This is the case for positive values, but the individual model output is higher compared to the mean model solar wind delay.
Table \ref{mean_values} shows the mean values of all features obtained from the  database.
For example, the mean speed of the interplanetary shocks detected at ACE is -469\,km/s in X-direction.
The trained random forest predicts a solar wind propagation delay of 47\,min when these mean feature values are used for input to the random forest.
The following scatter plots contain Shapley values for all 380 individual samples in the database.

\begin{table}[htbp]
  \centering
  \caption{Mean model input values (not standardized) and mean propagation delay from the fully trained Random Forest model.}\label{mean_values}
  \begin{tabular}{lllllll}
    \toprule
    $\overline{v_X}$\,[km/s]& $\overline{v_Y}$\,[km/s]&$\overline{v_Z}$\,[km/s]& $\overline{r_X}$\,[RE] & $\overline{r_Y}$\,[RE]& $\overline{r_Z}$\,[RE] &  $\overline{T_{delay}}\,[min]$ \\\midrule
    -469 & -14 & 0.6 & 233 & 0.89 & 0.29 & 47 \\\bottomrule
    \hline
  \end{tabular}
\end{table}

\begin{figure}[htbp]
  \centering
  \includegraphics[width=0.9\textwidth]{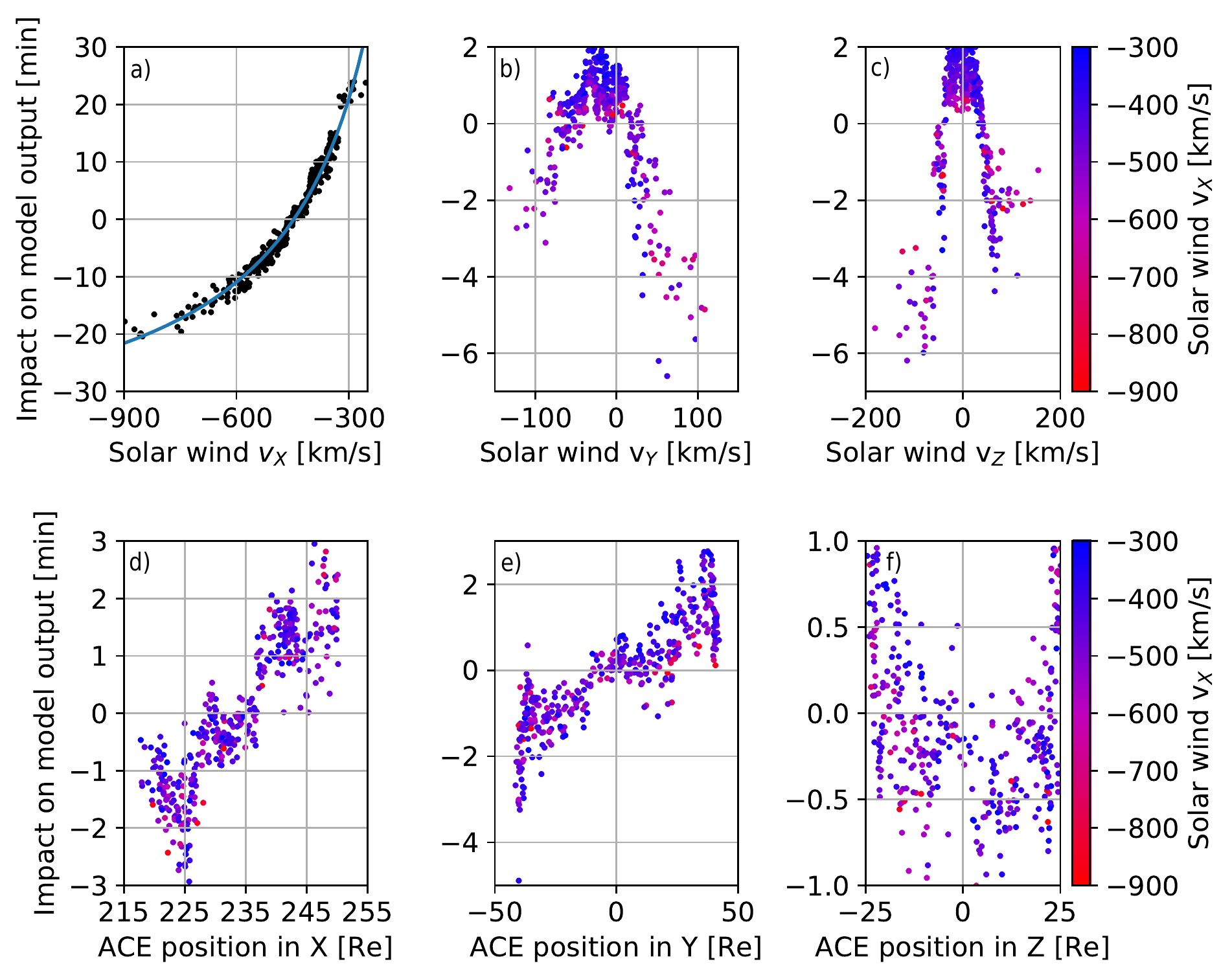}
  \caption{Shapley values for all six ML features, a) solar wind speed in X-direction, b) solar wind speed Y-direction, c) solar wind speed Z-direction, d) ACE position in X-direction, e) ACE position in Y-direction, f) ACE position in Z-direction, color-coded is the solar wind speed in X-direction.}\label{shap_speed}
\end{figure}

Figure \ref{shap_speed} panel a) shows the Shapley values for solar wind speed in X-direction, $v_x$, which has the biggest feature importance within the GB model.
The Shapley value reaches +/-20\,min for cases with $v_x =$ -300\,km/s and -900\,km/s respectively.
The relationship between $v_x$ and the Shapley values follows the assumption of constant solar wind speed and can be described by the function $t(v)=r/v - t_0$.
Here, $t_0$ can be identified as the mean solar wind delay and $r$ as the distance between ACE and the magnetopause.
The blue line in Figure \ref{shap_speed} a) indicates the best fit to the distribution of Shapley values.  The fit yields values for $t_0$= 43\,min and $r$= 198\,R$_E$ which are close to the actual mean values in Tab. \ref{mean_values}.

The Shapley value for solar wind speed in Y ($v_y$) and Z-direction ($v_z$) look completely different, see Fig. \ref{shap_speed} b), c).
An important difference to $v_x$ is that  $v_y$and $v_z$ vary between positive and negative values.
In both cases the relationship between shapley value and $v_y$/$v_z$ seem to have a quadratic form.
The Shapley value are negative for cases with high absolute solar wind speeds in Y and Z-direction, they can reach down to -6\,min.
Slightly positive Shapley values ($<$\,2 min) group around  $v_y$/$v_z$ being close to zero.
The shapley values for $v_y$ are shifted to negative solar wind speeds, i.e. close to the mean value of $v_y$ of -14 km/s.
For $v_z$, the parabola is closely centered around zero solar wind speed in Z-direction, as is the mean solar wind speed in that direction.

Shapley values for the position of ACE ($r_x$,$r_y$,$r_z$) are shown  in Figure \ref{shap_speed} d) ,e) ,f).
As ACE orbits around L1, it moves from 215 to 250\,R$_E$ in $r_x$, within +/- 50\,R$_E$ in $r_y$, and within +/-25\,R$_E$ in $r_z$ (see also Fig. \ref{learnset}).
The relationship between $r_x$ and the resulting Shapley values is linear and shows up to +/- 1\,min impact on model output for higher/smaller distances from Earth.
A similar linear relationship is exhibited by $r_y$ and its Shapley values, here negative $r_y$ values correspond to negative Shapley values of up to -2\,min and vice versa.
However, this linearity is affected by the solar wind speed in X-direction, $v_x$.
In the case of high solar wind speed, the Shapley values for $r_y$ are below $\pm2$\,min.
When  $v_x$ is rather small, the model output is more affected by $r_y$, especially when ACE is far from the Sun-Earh line.
For $r_z$ no linear relationship to the derived Shapley can be seen.
As already seen by the drop-column FI, also the Shapley value for $r_z$ are not much higher than $\pm$0.5 min.
When the ACE satellite is off the Sun-Earth line, the Shapley value reaches higher absolute values but the Shapley value can be both negative and positive.

\section{Discussion}\label{discussion}
The results described in Sec. \ref{results} show the possibilities of machine learning for the modeling of solar wind propagation delays. Here, we discuss and interpret the scientific impact of these results.

This study shows the possibility to time the solar wind propagation delay between a solar monitor at L1 and the magnetosphere. We used the magnetosphere's ability to act as a detector of interplanetary shocks for the timing of the solar wind propagation delay.  Precise timing is possible with the help of magnetometers not only onboard of satellites but also on Earth's surface.

It has to be noted here that  setting the target location rigidly to (15,0,0)\,R$_E$ introduces an additional error to the physical models. The time delays used in this study are however closely related to the interaction of the solar wind with the magnetosphere near the magnetopause. The location of the magnetopause varies between 6 and 15\,R$_E$ \citep[e.g.][]{Sibeck1991} and setting a fixed target location introduces an error in the order of one minute for the physical models. A bias has been identified in the mean error of the vector (2.8\,min) and flat delay method (5.2\,min) (see Fig. \ref{comparison} c)), which is influenced by the target location as well. Setting the target location to 10\,R$_E$ would have further increased that bias.

However, the comparison  between physics based modeling of SW propagation delay and trained ML algorithms remains fair because the ML algorithms do not have specific information about the location of the magnetopause either.  \citet{Cash2016} even use 30 Earth radii for their target to determine the solar wind delay, to account for property changes of the solar wind as it approaches the bow shock.

The cross validation shows a better performance of the trained gradient boosting model compared to the vector delay method. However, it has to be noted that the representation of the vector delay in this study might not be optimal. There are many parameters, e.g., the underlying technique to evaluate the normal vector or the number of data points used to define upstream and downstream conditions of the interplanetary shock, that govern the performance of the vector delay. A very detailed optimization of these parameters might further increase the performance of the vector delay but this is not the scope of this manuscript.

The good performance of the GB model is a favorable result by its own, but its  biggest advantage is the versatility of the ML approach. While, the vector delay method depends on a rigorous analysis of solar wind data during shock events, the ML approach only needs values for its six features at a single point in time to output a solar wind delay.  The training database consists of interplanetary shock events detected at ACE only, however the trained model can predict the arrival time of solar wind features of any kind at any given time. That allows for its application in a near real-time warning service for users in the space industries.

As the database used in this study relies on the timing of CMEs and CIRs only, a full generalization of the ML approach to times without interplanetary shocks remains to be investigated.
\citet{Cash2016} examined the generalization of the vector delay based on the minimum variance technique to a real time application, this study should be used as a blue print for the investigation on the ML approach. It has to be noted that the ML approach has deficiencies (see Fig.\,\ref{Flat_ML comparison}) when the input data is outside of the parameterspace used for training. This finding is in line with \citet{Smith2020}, who showed that ML based classification of sudden commencements can face misclassifications when applied outside of the trained parameter space.

A thorough analysis of the trained GB model improves the confidence in its solar wind propagation delay output.
That analysis includes an operational validation analysis with various train / test splits, a drop column feature importance analysis and a Shapley value analysis.

In the case of the operational validation analysis (Fig.~\ref{realtime_rmse}), GB performs better than both vector and flat methods when choosing an 80/20 or 90/10 split of the training and validation sets. Even though the vector method performs relatively well in its prediction, with sufficient training ( more than 200 cases of the database), the GB model achieves greater performance overall and has the potential for greater improvement if more training data instances  were available.

The importance of the six features of the dataset has been analysed based on a drop-column feature analysis. All features contribute to the performance of the solar wind propagation delay. The most important feature is the solar wind speed in X-direction, what is also expected from this problem.

Furthermore, the Shapley value analysis of the trained GB algorithm also gives additional confidence in the prediction of the solar wind propagation between L1 solar monitors and Earths magnetosphere.
From Fig. \ref{shap_speed} a,b,c) one can summarize that high solar wind speeds in any direction lead to a shorter propagation delay output from the GB model.
From Fig. \ref{shap_speed} d) it is obvious that a shorter distance between satellite and Earth corresponds to a shorter propagation delay output.

Slightly different is the case of the Y-component of ACE position around L1. From Fig.  \ref{shap_speed} e) it is obvious that the Y-component of the ACE position can increase, as well as decrease the modelled propagation delay. The decrease of propagation delay for negative values of $r_y$ and an increase for positive values can be accounted to the Parker spiral nature of the solar wind. Especially low speed cases show this effect.
High solar wind speed cases are less affected by the Parker spiral effect on the solar wind propagation delay.
A similar finding was shown by \citet[][Fig.4]{Mailyan2008} in their flat delay analysis, which shows a linear dependence of their flat delay error on the difference between the Y-component of the position of ACE and Cluster position.
The mean solar wind speed in the Y-direction of all cases in the database is also negative ( -14 km/s cf. Tab. \ref{mean_values}). This shift and its representation in the Shapley values of Fig. \ref{shap_speed} b) can be interpreted as an effect of the Parker spiral nature of the solar wind as well.

The DSCOVR satellite has been in L1 orbit since 2015 and will be the only solar monitor after the decommissioning of ACE and Wind. As the orbits of DSCOVR and ACE are very similar, we expect that our trained algorithm is also capable of predicting the solar wind propagation delay from DSCOVR data with similar accuracy. However, this has to be validated and is not within the scope of this manuscript.

Real-time predictions of SW propagation delay needs NOAA's real-time solar wind (RTSW) data. The results shown Fig.\,\ref{Flat_ML comparison} can be seen as the first successful demonstration of the continuous application of the ML approach.  However, the provided data is of different nature compared to the Level\,2 ACE data used in this study. The ACE RTSW data provides only bulk solar wind speed, proton density and three components of the magnetic field strength. In addition to that, RTSW data suffers from higher noise levels and additional data gaps. These problems impact the results of SW propagation delay predictions for ML as well as physical models. A future study will construct a new database based on RTSW data and investigate if a trained ML algorithms on that RTSW data can outperform the simple flat delay method. Additionally, the ML model could also benefit from information on the location of the magnetopause prior to a CME impact.

A future study can also investigate the role of the magnetic field before and after a interplanetary shock event to improve the predictions. Solar wind information upstream and downstream of  a shock can be used as additional ML features. By doing so, it can be investigated if giving additional information on the shock's normal vector could further improve the machine learning performance and provide a fairer comparison to the vector delay methods.
Furthermore, using the solar wind pressure may also be a feature to be included into the ML approach for better predictions, since it would help to indicate the position of the magnetopause.

The newly introduced ML approach to predict solar wind propagation delays from L1 data can be put into the group of velocity-based approaches like the flat and vector delay. A key drawback of simple velocity-based delay methods is that SW properties propagated to the target position can arrive out of order. There are schemes that already exist that address the problem of an unphysical propagation structure (e.g., OMNI, \citet{Weimer2008}).
 Hydrodynamic models on the other hand, which generate continuous data outputs based on the physical evolution of the plasma structure at L1, do not suffer from this drawback.

\section{Conclusions}\label{conclusions}

This work shows the possibility to model the solar wind propagation delay between L1 and Earth on the basis of machine learning. A database has been generated on the basis of ACE data and ground based magnetometer data which served as the training set for the random forest ML algorithm.
This database contains 380 measurements of the solar wind propagation delay from the detection of interplanetary shocks at ACE and their signature as sudden commencement in Earth's magnetosphere.

Random forest, gradient boosting and linear regression  have been applied to identify a suitable model for the SW propagation delay. Here, the gradient boosting algorithm performs best (RMSE = 4.5\,min), closely followed by the random forest and with larger margin, the linear regression. We also performed a hyperparameter optimization and found a slight improvement of 5-8\,\% to the default ML algorithms.

We performed a comparison of the ML model to physical models to derive the solar wind propagation delay, i.e., flat delay and vector delay.
The trained GB algorithm performs significantly better than the flat propagation delay model, i.e., the RMSE for the flat delay is more than 2\,min larger than for the GB approach.
The comparison showed that the vector delay method performs slightly worse compared to the trained GB model with an RMSE of 5.3\,min. An application of the GB model to continuous  solar wind data revealed that the predictions follow the flat delay as long as the solar wind speed in y and z is close to zero. The GB predictions give a shorter propagation time than the flat delay when that it is not the case. In addition, this analysis revealed that the GB output is closely related to the underlying training data. When operated outside the trained parameter space, e.g. when the solar wind is below 300 km/s, the GB model gives out unrealistic propagation delays.

The analysis of the trained GB algorithm was performed on the basis of a performance validation, feature importances and Shapley values.
The performance validation revealed that the GB model needs to be trained with at least 200 cases of the database in order to perform at par with the vector delay method.
In addition, the feature importance and Shapley value analysis enhanced the confidence in the trained GB algorithms and its predictions. The solar wind speed in X-direction was identified as the most important feature in the feature set. The Shapley value analysis revealed the internal relationship between the features and also indicated that the trained GB model follow basic physical principles like an empirical model.

The trained GB algorithm is suited to be run for post-event analysis or with near real-time data. The trained algorithm only needs input for solar wind speed vector and ACE's position vector to predict the solar wind propagation delay reliably. 

\begin{acknowledgements}
The authors thank Dmytro Vasylyev and Leonie Pick for valuable discussions on the topics of machine learning and Earth's magnetic field.

This paper uses data from the Heliospheric Shock Database (\url{ipshocks.fi}), generated and maintained at the University of Helsinki.
We acknowledge the provision of ACE data by the ACE science Center at Caltech (\url{www.srl.caltech.edu/ACE}).
The magnetometer data used in this work was made available by the INTERMAGNET consortium (\url{intermagnet.org}). We also acknowledge the SuperMAG service (\url{supermag.jhuapl.edu}) for magnetometer data visualisation.
This research made use of HelioPy, a community-developed Python package for space physics \citep{Stansby2019}.

The authors would like to thank two anonymous referees and Enrico Camporeale for their constructive comments which have helped to improve the manuscript.
\end{acknowledgements}

%
 \bibliography{ML_propa_delay}

%

\end{document}